\documentclass{amsart}
\usepackage{cite} \usepackage{url}
\usepackage{amsmath,amsthm,amssymb}

\newcommand\bF{\mathbb F} \newcommand\bP{\mathbb P} 
 \newcommand{\bbsm}{\left (\begin{smallmatrix}}      \newcommand{\besm}{\end{smallmatrix}\right )}
\newcommand\beq{\begin{equation}}  \newcommand\eeq{\end{equation}}
\newcommand\beqs{\begin{equation*}}  \newcommand\eeqs{\end{equation*}}
\newcommand\bep{\begin{proof}}  \newcommand\eep{\end{proof}}

\newcommand\omg{\omega}  \newcommand\mC{\mathcal C}
 \newcommand\mH{\mathcal H} \newcommand\mP{\mathcal P}
\newcommand\wkvd{\wedge^k V^*} \newcommand\wkv{\wedge^k V}

\newtheorem{theorem}{Theorem}[section]
\newtheorem{lemma}[theorem]{Lemma}
\newtheorem{proposition}[theorem]{Proposition}
\newtheorem{corollary}[theorem]{Corollary}
\newtheorem{definition}[theorem]{Definition}
\newtheorem{fact}[theorem]{Fact}

\begin{document}
\title{An asymptotic formula in $q$ for the number of $[n,k]$ $q$-ary MDS codes}
\author{Krishna~Kaipa}
      \thanks{K. Kaipa is with the department of Mathematics at the  Indian Institute of Science and Education Research, Bhopal  462066 India (email: kaipa@iiserb.ac.in). 
He is supported by the Indo-Russian project INT/RFBR/P-114 from the Department of Science \& Technology, Govt. of India.}

\begin{abstract}
We obtain an asymptotic formula in $q$ for the number of  MDS codes of length $n$ and dimension $k$ over a finite field with $q$ elements.
\end{abstract}

\maketitle

\section{Introduction}

MDS codes of dimension $k$ and length $n$ over $\bF_q$ are codes for which the minimum distance equals the Singleton bound $n-k+1$. Equivalently these are codes for which any $k \times n$ generator matrix has the property that all its $k \times k$ 
submatrices are nonsingular. Thus  $[n,k]_q$ MDS codes can be identified with 
the set of row equivalence classes of $k \times n$ matrices over $\bF_q$ having the property that any $k$ columns are linearly independent. Let $\gamma(k,n)$ denote the number of $q$-ary $[n,k]$ MDS codes. It is well known that the dual of an MDS code is also MDS, therefore $\gamma(k,n) = \gamma(n-k,n)$.  Consequently, we henceforth assume  without loss of generality that $k \leq n/2$. Exact values of $\gamma(k,n)$  are hard to determine and are known  only for $k=1,2$ all $n$, and for $k=3$ and $n \leq 9$. The formula for $\gamma(3,9)$ was found in the work  \cite{Sko} in 1995. Formulae for $\gamma(3,10)$ and  $\gamma(4,8)$ for example, are still unknown.  The known exact formulae can be written in the asymptotic form:
\beq \label{eq:asymp}
 \gamma(k,n) = q^{\delta} + \left(1 - N \right) \, q^{\delta-1} + a_2 \, q^{\delta-2}+ O(q^{\delta-3})
\eeq
where \beq \label{eq:Ndef} \delta: = k(n-k) \quad \text{and} \quad N: = \textstyle \binom{n}{k}. \eeq
We reproduce from the known exact formulae for $\gamma(k,n)$ (see \cite[\S 1]{Sko}) the corresponding values of $a_2(k,n)$.
\begin{eqnarray} \label{eq:known}
\nonumber a_2(1,n)&=&\frac{n^2 - 3 n +2}{2}\\ \nonumber 
a_2(2,n)& =&\frac{3 n^4 - 10 n^3 + 9 n^2  - 26 n +48}{24} \\
a_2(3,6) &=& 152\\ \nonumber
a_2(3,7) &=& 506 \\ \nonumber
a_2(3,8) &=& 1360 \\ \nonumber
a_2(3,9) &= & 3158
\end{eqnarray}

For general $(k,n)$, the following asymptotic formula was obtained in \cite{GL1}:
\beq \label{eq:GL1}  \gamma(k,n) = q^{\delta} + \left(1 - N \right) \, q^{\delta-1} + O(q^{\delta-2}) \eeq

Our main result is:
\begin{theorem} \label{asymp} The asymptotic formula \eqref{eq:asymp} holds for all $k$ and $n$, with 
\begin{eqnarray} \label{eq:unknown}
\nonumber a_2 &=& N \, k(n-k) \left(\frac{k^2 - nk+n+3}{2(k+1)(n-k+1)}\right) +\, N^2/2 - 5N/2 +2
\end{eqnarray}
\end{theorem}
The formulae  in  \eqref{eq:known} agree with the theorem.  The quantity 
\[ \tilde \gamma(k,n):=\gamma(k,n)/(q-1)^{n-1}\] is always an integer, as it is  the number of $n$-arcs in $PG(k-1;q)$ (see \cite[pp.219-220]{Sko1}, \cite[Lemma 2]{Sko}). We rewrite the theorem in terms of $\tilde \gamma(k,n)$: 

\begin{corollary} The number $\tilde \gamma(k,n)$ of $n$-arcs in $PG(k-1,q)$  is of the asymptotic form 
\[q^{\delta-n+1}  -  b_1 \, q^{\delta - n+2} + b_2 \,q^{\delta - n+3} +O(q^{\delta - n+4})\]
where $b_1 = N-n$, and
\[b_2 = a_2 - (n-1)(N-n) - (n^2 - 3n +2)/2\]
In particular:
\begin{eqnarray} \label{eq:unknown_crlry}
\nonumber \tilde \gamma(3,10) &=& q^{12} - 110 \,q^{11} + 5561 \,q^{10} +O(q^9) \\ 
\tilde \gamma(4,8) &=& q^9 - 62 \,q^8 + 1710 \,q^7 +O(q^6) 
\end{eqnarray}
\end{corollary}

The ideas of the proof of Theorem \ref{asymp} are as follows. Following \cite{Sko1, GL1}, we identify the set of MDS codes with a subset $U(k,n)$ of the Grassmannian $G(k,n)$ of $k$-dimensional subspaces of $\bF_q^n$. The subset $U(k,n)$ is an intersection of $N$ Schubert cells of $G(k,n)$, and hence its cardinality $|U(k,n)|$ can be expressed using inclusion-exclusion principle as a sum \[ E_1 - E_2 +E_3- \dots +(-1)^{N-1} E_N .\]
For each $r$, the quantity  $E_r$ is a sum of $\binom{N}{r}$ terms each of which is of the form $|G(k,n) \setminus L|$ where $L$ is a codimension $r$ linear subspace of the Pl\"{u}cker space $\bP(\wedge^k \bF_q^n)$ (in which $G(k,n)$ embeds). In \cite{GL1} asymptotic formulae of the form  $q^{\delta} + a q^{\delta-1} +O(q^{\delta-2})$ were obtained for each $|G(k,n) \setminus L|$, and hence for the $E_r$'s to obtain the result given in equation \eqref{eq:GL1}. We study linear sections $|G(k,n) \cap  L|$ more closely in section \ref{linear_sec}, leading to Theorem \ref{thm_codim_r} which gives an asymptotic formula of the form $q^{\delta} + a q^{\delta-1} + b q^{\delta-2} +O(q^{\delta-3})$ for $|G(k,n) \setminus L|$. These formulae in turn imply the desired asymptotic formulae for the $E_r$'s and hence for $\gamma(k,n)$.
\section{MDS codes as a subset of the Grassmannian} \label{sec2}
Let $V = \bF_q^n$ with standard basis $\{e_1, \dots, e_n\}$.
Let \beq \label{eq:multiset} I_{k,n}: = \{(i_1, i_2, \dots, i_k) : 1 \leq i_1 < \dots <i_k \leq n \}, \eeq
and for a multi-index $I = (i_1, \dots, i_k) \in I_{k,n}$ let  $e_I:=e_{i_1} \wedge \dots \wedge e_{i_k} \in \wedge^k V$. The multivectors $\{e_I : I \in I_{k,n} \}$ form the standard basis for $\wedge^k V$. Let $G(k,n)$ or $G_k(V)$ denote the Grassmannian  of $k$ dimensional  subspaces of $\bF_q^n$. We recall the  the Pl\"{u}cker embedding of  $G_k(V)$ into $\bP(\wedge^k V)$. For any $\Lambda \in G_k(V)$ and $k \times n$ matrix $M$ whose rows $b_1, \dots, b_k$ form a basis for  $\Lambda$, let  $i(\Lambda) := [b_1 \wedge \dots\wedge b_k] \in \bP(\wedge^k V) $. It is easy to see  that  $i(\Lambda)$ depends only on the row equivalence class of the matrix $M$, and hence only on $\Lambda$. Therefore we have a  function $i: G_k(V) \to \bP(\wedge^k V)$ which can also be shown to be injective. The image $i(G_k(V))$, which we will denote again by $G_k(V)$ is cut out by the quadratic  Pl\"{u}cker relations  \eqref{eq:plu}, and hence 
$i:G_k(V) \hookrightarrow \bP(\wedge^k V)$ realizes $G_k(V)$  as a projective variety. Expanding the expression $b_1 \wedge \dots\wedge b_k$ above in terms of the standard basis of $\wedge^k V$ we obtain 
\[b_1 \wedge \dots\wedge b_k = \sum_{I \in I_{k,n}} p_I(\Lambda) \, e_I\]
where $p_I(\Lambda)$ is the minor of the matrix $M$ on the columns indexed by $I$. It follows that $p_I(\Lambda), I \in I_{k,n}$ regarded as homogeneous coordinates, depend only on $\Lambda$ and are known as its  Pl\"{u}cker coordinates.\\

 Any $q$-ary code of dimension $k$ and length $n$, being a $k$ dimensional subspace of $\bF_q^n$, can be regarded as  point $\Lambda$ of $G_k(V)$. The matrix $M$ in the above discussion is precisely a generator matrix of the code $\Lambda$. As observed in the introduction, the code $\Lambda$ is $[n,k]$-MDS, if and only if all $k \times k$ submatrices of $M$ are non-singular. Hence the set of $[n,k]_q$ MDS codes can be identified with the subset of $G_k(V)$ defined as:
\beq \label{eq:ukn} U(k,n): = \{\Lambda \in G(k,n) : p_I(\Lambda) \neq 0, \; \forall \, I \in I_{k,n}
\}\eeq
In particular we note
\[ \gamma(k,n) = |U(k,n)| \]
For each of the $N = \binom{n}{k}$ multi-indices $I \in I_{k,n}$ let
\beq \label{eq:cell} C_I: =  \{\Lambda \in G(k,n) : p_I(\Lambda) \neq 0\}\eeq 
For each $\Lambda \in C_I$, there is a unique $k \times n$ matrix $M$ whose rows forms a basis for $\Lambda$, and which satisfies the property that  the $k\times k$ submatrix of $M$ on the columns indexed by $I$ is the $k \times k$ identity matrix.
Conversely the row space of a matrix $M$ which has the $k \times k$ identity matrix as the submatrix on the columns indexed by $I$,  is a point of $U(k,n)$. Therefore
$C_I$ can be identified with the set of all such matrices. Since the columns of $M$ indexed by $I$ are fixed and the remaining $n-k$ columns free, we see that 
\beq \label{eq:CI} |C_I| = q^{k(n-k)} = q^{\delta}. \eeq \\

It follows from the definitions \eqref{eq:ukn} and \eqref{eq:cell}, that
\beq \label{eq:ukncell} U(k,n) = \bigcap_{I \in I_{k,n}} C_I \eeq
For each $r$ with $1\leq r \leq N$, let  $I^r_{k,n}$ denote the subsets of $I_{k,n}$ of cardinality $r$. Let 
\beq \label{eq:E_r} E_r: = \sum_{\{I_1, \dots, I_r\} \in I^r_{k,n}} | C_{I_1} \cup \dots \cup C_{I_r}| \eeq
It follows from the inclusion-exclusion principle applied to \eqref{eq:ukncell} that
\beq \label{eq:ukncard} 
|U(k,n)| = E_1 - E_2 +E_3- \dots +(-1)^{N-1} E_N
 \eeq
The quantity $E_r$ is a sum of $\binom{N}{r}$ terms of the form $|C_{I_1} \cup \dots \cup C_{I_r}|$. The set $G_k(V) \setminus  ( C_{I_1} \cup \dots \cup C_{I_r})$ consists of points of $G_k(V)$ having Pl\"{u}cker coordinates $p_{I_1}, \dots, p_{I_r}$ zero. Therefore if  $\hat L$ is the codimension $r$ subspace of $\wedge^k V$ spanned by $\{e_I : I \notin \{I_1, \dots, I_r\}\}$,  and  $L =\bP \hat L$, then $G_k(V) \setminus  ( C_{I_1} \cup \dots \cup C_{I_r}) = G_k(V) \cap L$.  The set $G_k(V) \cap L$ is an example of a linear section of $G_k(V)$ which we study in the next section.

\section{Linear sections of the Grassmannian} \label{linear_sec}
We say $L \subset \bP(\wedge^k V)$ is a codimension $r$ linear subspace if  $L = \bP \hat L$  with $\hat L$ is a codimension $r$ linear subspace of $\wedge^k V$.
Corresponding to the standard basis $e_1, \dots, e_n$ of $V$, the dual space $V^*$ has the standard dual basis $e^1, \dots, e^n$ defined by $\langle e^i, e_j \rangle  = \delta_{ij}$.
Here $\langle, \rangle$ is the natural pairing between $V^*$ and $V$. Similarly the space $\wedge^k V^*$ has the standard basis $\{e^I : I \in I_{k,n}\}$  where $e^I=e^{i_1} \wedge \dots \wedge e^{i_k}$. Given $k$ elements $\omega_1, \dots, \omega_k$ of $V^*$ and $k$ elements $v_1, \dots, v_k$ of $V$, the determinant of the $k \times k$ matrix with entry in row $i$  and column $j$ being $\langle \omega_i, v_j \rangle$, is multilinear and alternating in both $(\omega_1, \dots, \omega_k)$ and  $(v_1, \dots, v_k)$. It therefore defines a bilinear pairing $\langle, \rangle$ between $\wedge^k V^*$ and $\wedge^k V$. In terms of the standard bases we have $\langle e^I, e_J\rangle = \delta_{IJ}$, which also shows that the pairing is non-degenerate and hence gives an isomorphism between $\wedge^k V^*$ and $(\wedge^k V)^*$. We refer to elements of $\wedge^k V^*$ in short as $k$-forms.

Given a codimension $r$ linear subspace $L=\bP(\hat L)$  of $\bP(\wedge^k V)$, let \[\text{Ann}(\hat L): = \{ \omega \in \wedge^k V^* : \langle \omega,\xi \rangle = 0, \; \forall \, \xi \in \hat L\},\]
 and let Ann$(L) = \bP (\text{Ann}(\hat L))$. We note that Ann$(L)$ is  a $r-1$ dimensional linear subspace of $\bP(\wedge^k V^*)$. The correspondence $L \leftrightarrow \text{Ann}(L)$ allows us to identify codimension $r$ linear subspaces of $\bP(\wedge^k V)$ with $r-1$ dimensional linear subspaces of $\bP(\wedge^k V^*)$.
 We define
\beq \label{eq:normL}  ||L|| := |G_k(V) \setminus L| \eeq
We call a codimension $1$ subspace $H$ of $\bP(\wedge^k V)$, a hyperplane. It follows from the above discussion that each hyperplane corresponds to a unique $k$-form (upto a nonzero scalar multiple) $\omega$. We write $H = H_{\omega}$ to emphasize this, and we also define 
\beq \label{eq:normH} ||\omega|| := ||H_{\omega}||  = |G_k(V) \setminus H_{\omg}| \eeq

The main results of this section are Theorems \ref{wt2_thm} and \ref{thm_codim_r} which appear at the end. In the remaining part of this section, we either recall or develop several results leading upto the main results.

\begin{definition} A linear subspace of $G_k(V)$ is a  linear subspace of $\bP(\wkv)$ which is entirely contained in $G_k(V)$. 
\end{definition} 
 For each $\alpha \in G_{k-1}(V)$ and each $\gamma \in G_{k+1}(V)$, we define:
\begin{eqnarray}
 \pi_{\alpha} &:=& \{\beta \in G_k(V) : \beta \supset \alpha\}\\ \nonumber
\pi^{\gamma} &:=& \{ \beta \in G_k(V) : \beta \subset \gamma\}
\end{eqnarray}

Since $\pi_{\alpha}$ is projectively isomorphic to $\bP(V/\alpha)$, it is a linear subspace of $G_k(V)$ of dimension $n-k$. Similarly  $\pi^{\gamma}$ is projectively isomorphic to $\bP(\gamma^*)$, and hence it is a linear subspace of $G_k(V)$ of dimension $k$.  It is a classical fact \cite[\S 2.1]{Chow}, \cite[Proposition 3.2]{Pankov} that, for $k \geq 2$, the $\pi_{\alpha}$'s and the $\pi^{\gamma}$'s are the  maximal linear subspaces of $G_k(V)$. Some facts about these spaces that we will need are as follows. They easily follow from the definitions of $\pi_{\alpha}$ and $\pi^{\gamma}$ (also see \cite[p.88]{GK}).
\begin{fact} \label{fact0} \begin{enumerate} \item For $\alpha \neq \alpha' \in G_{k-1}(V)$ the intersection
$\pi_{\alpha} \cap \pi_{\alpha'}$ is empty if dim$(\alpha + \alpha')  > k$ and consists of the single point $\alpha+\alpha'$ if dim$(\alpha +\alpha') = k$. 
\item  For  $\gamma \neq \gamma'  \in G_{k+1}(V)$ the intersection $\pi^{\gamma} \cap \pi^{\gamma'}$ is empty if dim$(\gamma \cap \gamma') < k$ and consists of the single point 
$\gamma \cap \gamma'$ is dim$(\gamma \cap \gamma') =k$.
\item The intersection $\pi_{\alpha} \cap \pi^{\gamma}$ is empty if $\alpha \not\subset \gamma$, and it is the  line $\{\beta: \alpha \subset \beta \subset \gamma\}$ if $\alpha \subset \gamma$. 
\item For any pair $\beta, \beta'  \in \pi_{\alpha}$ the intersection $\beta \cap \beta'$ is $\alpha$, and for any pair $\beta, \beta'  \in \pi^{\gamma}$ the vector space sum $\beta + \beta'$ is $\gamma$. \end{enumerate} \end{fact}

We recall the definition of the interior multiplication operator: for $\xi \in \wedge^{\ell} V$,  $\zeta \in \wedge^m V$, and $\omega \in \wedge^{\ell+m} V^*$,  $ \iota_{\xi} \omega \in \wedge^m V^*$ is defined by:
\[ \langle \iota_{\xi} \omega, \zeta \rangle  = \langle \omega, \xi \wedge \zeta \rangle \]
For any nonzero $\omg \in \wkvd$, we define subspaces $V_{\omg} \subset V$ and $U_{\omg} \subset V^*$ by 
\begin{eqnarray}
V_{\omg} &:=& \{v \in V :  \iota_v \omg = 0\} \\ \nonumber
U_{\omg}&:=& \{\theta \in V^* : \langle \theta, v \rangle=0 \quad \forall \, v \in V_{\omg}\}
\end{eqnarray}

\begin{fact} \label{fact1} We recall \cite[p.210]{GH} that $\omg$ is decomposable if and only if dim$(V_{\omg})= n-k$ or equivalently dim$(U_{\omg}) = k$.
We also note that if $\omg$ is indecomposable then dim$(U_{\omg}) \geq k+2$ because $\omg \in \wedge^k U_{\omg}$ and 
every element of $\wedge^k \bF^{k+1}$ is decomposable. \end{fact}
The next lemma will be used in the proof of Theorem \ref{wt2_thm}.
\begin{lemma} \label{dec_lemma} Let $k \geq 2$. A $k+1$-form $\omg \in \wedge^{k+1} V^*$  is decomposable if and only if $\iota_v \omg$ is decomposable for all $v \in V$. \end{lemma}
\bep If $\omg$ is decomposable, then so is $\iota_v \omg$. Conversely if $\iota_v \omg$ is decomposable for all $v \in V$, then 
\[ S_{\omg}:=\bP\{\iota_v \omg : v \in V\} \subset G_{k}(V^*) \subset \bP(\wedge^{k} V^*) \] is a linear subspace of  $G_{k}(V^*)$. 
Therefore $S_{\omg}$ is contained either in a $\pi_{\alpha}$ for some $\alpha \in G_{k-1}(V^*)$, or a  $\pi^{\gamma}$  for some $\gamma \in G_{k+1}(V^*)$.
Suppose $S_{\omg} \subset \pi_{\alpha}$.  Let $f_1, f_2, \dots, f_n$ be a basis of $V$ such that $f_1, \dots, f_{\ell}$ is a basis of $V_{\omg}$ or equivalently
$f^{\ell+1}, \dots,  f^n$ is a basis of $U_{\omg}$ where  $f^1, \dots, f^n$ is the dual basis.  
The condition $n-\ell = k+1$ for $\omg$ to be decomposable  can be written  as $n-\ell \leq k+1$ since $n- \ell$ equals dim$(U_{\omg}) \geq k+1$. The expression for $\omg$ 
in terms of the basis vectors $\{f^I  : I \in I_{k+1,n}\}$ of $\wedge^{k+1} V^*$ involves only those basic vectors $f^I$ for which all indices of $I$ are strictly greater than $\ell$.
For each $ i \geq \ell+1$, the fact that $\iota_{f_i} \omg \neq 0$ implies that there is at least one $j \neq i$ with $j  \geq \ell+1$ such that $\iota_{f_i \wedge f_j} \omg \neq 0$.
Now let  $\beta = \iota_{f_{i}} \omg$   and $\beta' =  \iota_{f_{j}} \omg$.
Then $\iota_{f_i \wedge f_j} \omg  = -\iota_{f_i} \beta' = \iota_{f_j} \beta$ represents a
$k-1$ dimensional space contained in the $k$ dimensional spaces $\beta$  and $\beta'$.
It follows that $\iota_{f_i \wedge f_j} \omg$ represents $\beta \cap \beta'$.
By part 4) of Fact \ref{fact0}, we conclude that the expression for $\alpha$ in terms of the basis vectors $\{f^I : I \in I_{k-1,n} \}$  of $\wedge^{k-1} V^*$ involves only those basis vectors $f^I$ for which no index of $I$ equals $i$. Since $i \geq \ell+1$ was arbitrary, the expression for $\alpha$ involves only those basic vectors $f^I$ for which all indices of $I$ are less than or equal to $\ell$. This however contradicts the fact that the expression for $\omg$ and hence $\iota_{f_i \wedge f_j} \omg$ involves only those $f^I$ for which the indices of $I$ are strictly greater than $\ell$. This contradiction shows that $S_{\omg} \subset \pi^{\gamma}$ for some $\gamma \in G_{k+1}(V^*)$. Since dim$(\pi^{\gamma}) = k$ and dim$(S_{\omg}) = n - \ell - 1$ we get  $n-\ell \leq k+1$  as desired. \eep

The next lemma will be used in the proof of Proposition \ref{codim1_prop}.
\begin{lemma} \label{pairwise}  If $\mP$ is a subset of $G_k(V)$ such that for every pair $P, Q \in \mP$, the line $\overline{PQ}$ joining them is contained in $G_k(V)$, then the linear subspace of  $\bP(\wedge^k V)$ generated by $\mP$ is completely contained in $G_k(V)$.
 \end{lemma}
\bep An element $\lambda \in \wedge^k V$ is decomposable if and only if it satisfies the Pl\"{u}cker relations (\cite[pp.210-211]{GH})
\beq \label{eq:plu} (\iota_{\xi} \lambda) \wedge \lambda  = 0 \quad \forall \, \xi \in \wedge^{k-1} V^*\eeq
Let $\lambda_1, \lambda_2, \dots,\lambda_n$ be decomposable elements of $\wedge^k V$ representing points $P_1, P_2, \dots, P_n$ of $\mP$. For each $\xi \in \wedge^{k-1} V^*$,  
let $a_i(\xi): = (\iota_{\xi} \lambda_i) \wedge \lambda_i$  for $1 \leq i \leq n$, 
and let  $a_{ij}(\xi): = (\iota_{\xi} \lambda_i) \wedge \lambda_j + (\iota_{\xi} \lambda_j) \wedge \lambda_i$ for  $1 \leq i < j  \leq n$.  The condition for $\lambda = \sum t_i \lambda_i$ to be decomposable is \[\sum_i t_i^2 \, a_i(\xi)  + \sum_{i<j} t_i t_j  \, a_{ij}(\xi)  = 0, \;  \forall \, \xi \in \wedge^{k-1} V^*\]
 Using \eqref{eq:plu}, $a_i(\xi) = 0$ because
$\lambda_i$ is decomposable, and $a_{ij}(\xi)=0$ because $\overline{P_i P_j} \subset G_k(V)$.
\eep
\medskip

Some of the results we will need about the cardinalities $||L||$ (as defined in \eqref{eq:normL}) appear in the literature in the context of linear codes associated to the embedding $G(k,n) \hookrightarrow \bP(\wedge^k \bF_q^n)$. We briefly describe the construction of these Grassmann codes $\mC(k,n)$, the details of which can be found in \cite{Nogin}. These are linear codes of length $\tilde n$ and dimension $\tilde k$ where:
\begin{eqnarray}
\tilde n = |G(k,n)| &=&\frac{(q^n-1) (q^{n-1}-1)\!\ldots \! (q^{n-k+1}-1)}{(q^{k}-1)(q^{k-1}-1)\!\ldots \!(q-1)} \nonumber \\
\tilde  k &=& \textstyle \binom{n}{k}
\end{eqnarray}

Let $P_1, P_2, \dots, P_{\tilde n}$ denote representatives in $\wedge^k \bF_q^n$ of the $\tilde n$ points of $G(k,n)$ arranged in some order. We also arrange in some order, the $\tilde k$ basic vectors $\{e_I : I \in I_{k,n}\}$ of $\wedge^k \bF_q^n$. 
A $\tilde k \times \tilde n$ generator matrix for the code $\mC(k,n)$ has for its $(i,j)$-th entry, the $i$-th Pl\"{u}cker coordinate $p_i(P_j)$. The fact that the Pl\"{u}cker embedding is non-degenerate (i.e no hyperplane of $\bP(\wedge^k \bF_q^n)$  contains $G(k,n)$) is equivalent to the fact that the generator matrix above has full row rank. A codeword of 
$\mC(k,n)$ is of the form $(\sum_I a_I p_I(P_1), \dots, \sum_I a_I p_I(P_{\tilde n}))$.
If $\omega$ is the $k$-form $\sum_I a_I e^I$, then the above codeword can be expressed as $(\omega(P_1), \dots, \omega(P_{\tilde n}))$. In this way we identify codewords of $\mC(k,n)$ with $k$-forms on $V = \bF_q^n$. Moreover, the Hamming weight of the above codeword clearly coincides with the expression $||\omega||$ as defined in  \eqref{eq:normH}. More generally, let $\hat L$ be a codimension $r$ subspace of $\wedge^k \bF_q^n$, and let $\{\omega_1, \dots, \omega_r\}$ be a basis for Ann$(\hat L)$. 
Evaluating elements of Ann$(\hat L)$, i.e forms $\sum_{i=1}^r a_i \omega_i$ on $P_1, \dots, P_{\tilde n}$ gives a $r$ dimensional subcode of $\mC(k,n)$. Moreover, the Hamming weight of this subcode is clearly  $||L||$ as defined in  \eqref{eq:normL} (where $L = \bP(\hat L)$). The minimum distance $d_1(\mC(k,n))$  of the code $\mC(k,n)$ equals 
 minimum value of $||\omega||$ over all nonzero $k$-forms on $V$, and the higher weight $d_r(\mC(k,n))$ equals the minimum value of $||L||$ over all codimension $r$  subspaces of $\bP(\wedge^k V)$. 

Using the well known formula (see \cite[p.7]{Tsfasman})
\[ ||D|| = \frac{1}{q^r - q^{r-1}} \sum_{c \in D} ||c||\]
 relating the Hamming weight of a $r$ dimensional subcode $D$ to the Hamming weights of its constituent codewords, we get the corresponding formula
\beq \label{eq:normLH}  || L || = \frac{1}{q^{r-1}} \sum_{[\omg] \in \text{Ann}(L)} || \omg||  =\frac{1}{q^{r-1}} \sum_{H  \supset L} ||H|| \eeq
where the sum $H \supset L$ is over all hyperplanes $H$ containing $L$.\\

The next theorem summarizes the information we will need about the minimum distance, higher weights and the weight spectrum of the code $\mC(k,n)$.
\begin{theorem} [Nogin]  \cite{Nogin} \label{thmNogin} \begin{enumerate}
\item The minimum distance $d_1(\mC(k,n))$ equals $q^{\delta}$. The codewords  of minimum weight correspond precisely with decomposable $k$-forms.

\item More generally, for $1 \leq r \leq n-k+1$, the higher weight \[ d_r(\mC(k,n)) =q^{\delta} +q^{\delta-1} + \dots+q^{\delta - r+1}.\]
 The $r$ dimensional subcodes with minimum weight correspond exactly to codimension $r$ subspaces $L$ of $\bP(\wedge^k V)$ such that  Ann$(L)$ is a linear subspace of $G_k(V^*)$.

\item For the code $\mC(2,n)$, upto a code-isomorphism any codeword corresponds to one of the  $2$-forms \[ \omega_r = e^1 \wedge e^2 + e^3 \wedge e^5 + \dots +e^{2r-1} \wedge e^{2r},\]
where $1 \leq r \leq \lfloor n/2 \rfloor$. Moreover \[||\omega_r|| = q^{\delta} +q^{\delta-2} + \dots+ q^{\delta - 2r+2} \] 
\end{enumerate}
\end{theorem}

\bep We refer to \cite{Nogin} for the proofs of parts 1) and 3). We give a quick proof of part 2) of the theorem. For a codimension $r$ subspace $L \subset \bP(\wedge^k V)$, there are $1+q+q^2+ \dots+q^{r-1}$ elements in Ann$(L)$, and each  $[\omega] \in \text{Ann}(L)$ satisfies $||\omega|| \geq q^{\delta}$ (by the first part of the theorem). The formula \eqref{eq:normLH} gives 
\begin{eqnarray}  \label{eq:normLbnd}
\nonumber ||L|| \geq \frac{q^{\delta} (1+q+ \dots+ q^{r-1})}{q^{r-1}}
= q^{\delta} +q^{\delta-1} + \dots+q^{\delta - r+1}
\end{eqnarray}
Moreover, equality holds above if and only if each $[\omega]$ in Ann$(L)$ satisfies $||\omega|| = q^{\delta}$,  or equivalently  is decomposable. In other words equality holds if and only if Ann$(L)$ is a $r-1$ dimensional linear subspace of $G_k(V^*)$.  Since linear subspaces of $G_k(V^*)$ exist only in dimensions less than or equal to $n-k$, taking such subspaces for Ann$(L)$ we obtain the desired formula for $d_r(\mC(k,n))$ when $1 \leq r \leq n-k+1$. \eep

The next lemma relates the weight of a $\mC(k,n)$ codeword to certain codewords of $\mC(k-1,n-1)$ associated with it. For a nonzero $\omg \in \wkvd$ and a vector $u \notin V_{\omg}$, let $\omg_{u}$ denote  $\iota_u \omg$  regarded as a $k-1$ form on the quotient vector space $V/(\bF_q u)$. The weight $||\omg||$ can be expressed in terms of the  weights $||\omg_u||$ for $u \notin V_{\omg}$ as follows.
\begin{lemma} \label{wt1_lm}
\[ ||\omg|| = \frac{1}{q^k-1}\, \sum_{u \notin V_{\omg}} ||\omega_{u}||\] \end{lemma}
\bep Let us denote the cardinality of the general linear group $GL(m,\bF_q)$ by
\[[m]_q := q^{m(m-1)/2} \,  (q^m-1)(q^{m-1}-1) \cdots (q-1)\]
Let $[v_1, v_2, \dots, v_{k}]$ denote an ordered set of $k$ vectors of $V$.
By definition of $||\omg|| = | G_k(V) \setminus \mH_{\omg}|$ we get:
\begin{eqnarray*}
 [k]_q \!\cdot ||\omega|| &=& \left| \{ [v_1, v_2, \dots, v_k] \, : \, \langle \omega, v_1 \wedge \dots \wedge v_k \rangle \neq 0 \} \right|  \\
&=&  \sum_{u \notin V_{\omg} } \left| \{ [v_2, \dots, v_k] \, : \,  \langle \iota_{u}\omega,  v_2 \wedge \dots \wedge v_k \rangle \neq 0 \} \right|.
\end{eqnarray*} 
Let $\tilde v_i$ for $2 \leq i \leq k$ denote the class of $v_i$ in $V/(\bF_q u)$.  There are $q^{k-1}$  ordered sets $[v_2, \dots, v_k]$ whose projection under $V \to V/(\bF_q u)$ 
is a given ordered set $[ \tilde v_2, \dots, \tilde v_k]$. 
Therefore,  
\beqs \frac{[k]_q  \!\cdot ||\omega||}{ q^{k-1} } = 
  \sum_{u \notin V_{\omg} } \left| \{ [\tilde v_2, \dots, \tilde v_k] \, : \,  \langle \omega_u,  \tilde v_2 \wedge \dots \wedge \tilde v_k \rangle \neq 0 \} \right| 
= [k-1]_q  \sum_{u \notin V_{\omg}} ||\omega_{u}||.
\eeqs

The lemma now follows by noting that \[ [k]_q =  q^{k-1} (q^k-1) \, [k-1]_q.\]
\eep \medskip

\begin{proposition} \label{codim1_prop}
 For $1 \leq m \leq n-k+1$,  let  $L$ be an $m$ dimensional subspace of $\bP(\wkv)$ such that $L$ is not contained in $G_k(V)$. Suppose $L_1 \subset L$ is an $m-1$ dimensional subspace which is contained in $G_k(V)$. Then we have:
\begin{equation} \label{eq:codim1}  |L \cap G_k(V)|   - |L_1| \leq  
\begin{cases}
1 & \text{if  } m=1, \\
 q  & \text{if  } m=2, \\
q^2  & \text{if  } m  \geq 3
\end{cases}
\end{equation}
\end{proposition}

\bep  If there is no point $P \in L \cap G_k(V)$ which is not in $L_1$, then $|L \cap G_k(V)| -|L_1|$ being zero, clearly satisfies the stated bounds. So we assume such a point $P$ exists. Let $L_2$ be the subset of $L_1$ defined by
\[ L_2 := \{Q \in L_1: \overline{PQ} \subset G_k(V) \}\] 
where, for any two points $P_1 \neq P_2 \in G_k(V)$, $\overline{P_1 P_2}$  denotes the line  joining $P_1$ and $P_2$. By extending a basis of $P_1 \cap P_2$ to a basis of $P_1$ and $P_2$, it is easy to see that
\beqs
\overline{P_1 P_2} \cap G_k(V) = \begin{cases} 
\overline{P_1 P_2}  & \text{if  } \;   \text{dim}(P_1 \cap P_2)=  k-1 \\
\{P_1, P_2\} & \text{if   }   \; \text{dim}(P_1 \cap P_2) <  k-1
\end{cases}
\eeqs
This together with the fact that $L = \cup_{P' \in L_1} \overline{PP'}$, gives:

\beq \label{eq:LL1L2}  |L \cap G_k(V)|  - |L_1| = 1+ (q-1) |L_2| \eeq

Let $\mP=L_2 \cup \{P\}$. Given $P_1, P_2 \in \mP$, if $P_1$ or $P_2$ equals $P$, then by definition of $L_2$ the line $\overline{P_1P_2}$ is contained in $G_k(V)$. Otherwise $P_1, P_2 \in L_2 \subset L_1$, and since $L_1 \subset G_k(V)$ is a linear subspace we again get
 $\overline{P_1P_2} \subset G_k(V)$. Therefore we can apply Lemma \ref{pairwise} to 
$\mP$ to conclude that the linear subspace $L(\mP)$ generated by $\mP$ is contained in $G_k(V)$. Let $\pi'$ be a maximal linear subspace of $G_k(V)$ containing $L(\mP
)$. If  $Q, Q' \in L_2$ and $Q'' \in \overline{QQ'}$, then $\overline{PQ''} \in L(\mP)$, and hence  $\overline{PQ''} \subset G_k(V)$. This shows that $Q'' \in L_2$, and hence that 
$L_2$ is a linear subspace of $G_k(V)$.

Let $\pi$ be a maximal linear subspace of $G_k(V)$ containing $L_1$. We note that $P \notin \pi$, for otherwise all lines joining $P$ to $L_1$ are contained in $G_k(V)$, and hence $L$ itself,  being the union of these lines, is contained in $G_k(V)$, which is not true by hypothesis.
Therefore $\pi \neq \pi'$. By parts 1)-3) of  Fact \ref{fact0}, $\pi \cap \pi'$ is either a line, a point  or empty. Since $L_2$ is a linear subspace  of $\pi \cap \pi'$,  we conclude that $|L_2| \in \{0,1,1+q\}$. Using this in \eqref{eq:LL1L2} we get: \[ |L \cap G_k(V)| - |L_1| \in \{1, q, q^2\}.\] The fact that $L \not\subset G_k(V)$ implies $|L \cap G_k(V)| - |L_1| < q^m$. Therefore, in the  case $m=1$ we get
 $|L \cap G_k(V)| - |L_1| =1$, and in the case $m=2$,  we get $|L \cap G_k(V)| - |L_1| \in \{1, q\}$. 
\eep
\medskip
\begin{proposition} \label{wt_prop}
Let $L$ be an $\ell$ dimensional linear subspace of $\bP(\wkv)$ which is not contained in $G_k(V)$. Also suppose $\ell \geq 3$. Then: 
\[ |L \cap G_k(V)| \leq 1+q +2 q^2+q^3 + \dots+q^{\ell-1}\]

\end{proposition}

\bep Let $L''$ be a  subspace of $L$ which is maximal with respect to the property of being contained in $G_k(V)$.  If $L''= \varnothing$ then $ |L \cap G_k(V)| = 0$ satisfies the asserted bound. So we assume dim$(L'') = \mu \geq 0$ and let
\[ \mathcal F := \{L' \subset L :  {\rm dim}(L') = \mu+1, L' \supset L''\}\]
We note that $|\mathcal F| = |\bP^{\ell-\mu-1}|$.
Each point of $L$ is contained in some $L' \in \mathcal F$, and 
any two distinct elements of $\mathcal F$ intersect in $L''$.
Therefore
\[ |L \cap G_k(V)| = |L''| + \sum_{L' \in \mathcal F} ( |L' \cap G_k(V)| - |L''|)\]
The pair of spaces $L'' \subset L'$ satisfies the hypothesis of Proposition \ref{codim1_prop}, therefore  $ |L' \cap G_k(V)| - |L''|$  satisfies the bounds of \eqref{eq:codim1}. Consequently,
\beqs
 \sum_{L' \in \mathcal F} ( |L' \cap G_k(V)| - |L''|) \leq
\begin{cases}
1+q+\dots+q^{\ell-1} & \text{if  } \mu=0, \\
 q+q^2+\dots+q^{\ell-1}  & \text{if  } \mu=1, \\
q^2+q^3+\dots+q^{\ell-\mu+1}  & \text{if  } \mu  \geq 2
\end{cases}
\eeqs
Adding $||L''|| = 1+q+\dots+q^{\mu}$ to the above equation we get:
\beq  \label{eq:mueq01} |L \cap G_k(V)|  \leq \\
\begin{cases}
2+q+ \dots+q^{\ell-1} &\!\!\text{if  } \mu=0 \\
 1+2q+q^2+\dots+q^{\ell-1}  &\!\!\text{if  } \mu=1 \end{cases}
\eeq
and if $\mu \geq 2$  then 
\begin{eqnarray}  \label{eq:mugeq2}
|L \cap G_k(V)|  &\leq&
1+q+2 \left(q^2+\dots+q^{\text{min}\{\mu, \ell-\mu+1\}} \right)\\ \nonumber
&& + \, \left( q^{\text{min}\{\mu+1, \ell-\mu+2\}} +\dots+q^{\text{max} \{\mu, \ell-\mu+1\}}\right) 
\end{eqnarray}
Consider the quantity:
\[ A:  = 1+q +2 q^2+q^3 + \dots+q^{\ell-1}\]
and let $B$ denote the right hand side of \eqref{eq:mueq01} if $\mu =0$ or $\mu=1$, and the right hand side of \eqref{eq:mugeq2} if $2 \leq \mu \leq \ell-1$.
The assertion in the proposition statement that needs to be established is  $|L \cap G_k(V)| \leq A$. Hence it suffices to show $A \geq B$. If $\mu =0$, then $A-B =q^2-1 >0$ and if $\mu=1$ then $A-B= q^2-q >0$. If $\mu = 2$ or $\ell - 1$, then $A=B$. In the remaining cases $3 \leq \mu \leq \ell-2$, we get:
\[\frac{(q-1)(A-B)}{q^3}= (q^{\ell-1-\mu} - 1) (q^{\mu-2} - 1)\]
Since $3 \leq \mu \leq \ell-2$ implies  $\ell- 1-\mu \geq 1$ and $\mu-2 \geq 1$, we obtain $A>B$.  
\eep
\begin{theorem} \label{wt2_thm} Let  $\omg \in \wkvd$ be a nonzero $k$-form.  If $\omg$ is  decomposable then $||\omg|| = q^{\delta}$, and if $\omega$ is indecomposable then 
\beq \label{eq:wtindec} ||\omg|| =  q^{\delta} +q^{\delta-2} + O(q^{\delta-3})  \eeq
(where $\delta(k,n) = k(n-k)$ and $k \leq n/2$) \end{theorem}
\bep If $\omg$ is decomposable then  $||\omg|| = q^{\delta}$ by 
part 1) of Theorem \ref{thmNogin}.  If $\omg$ is indecomposable and $k=2$, then the desired result \eqref{eq:wtindec} holds by part 3) of Theorem \ref{thmNogin}, for all $n\geq 4$.
 So, we  assume  $k \geq 3$ and  assume inductively that the result holds for all $k-1$ forms $\omg$ on a vector space of dimension $n \geq 2(k-1)$. We now use Lemma \ref{wt1_lm} and the notation therein. Let $W_{\omg}$ be a complement of $V_{\omg}$ in $V$, so that every element of $V \!\setminus\! V_{\omg}$ can be written as $u +v$ with $u \in W_{\omg} \!\setminus\! \{0\}$ and $v \in V_{\omg}$. 
We let dim$(W_{\omg}) = $ dim$(U_{\omg}) = k+s$ where $s \geq 2$  as noted in Fact \ref{fact1}. We get:
\beq \label{eq:wt2eq1} ||\omg|| = \frac{q^{n-k-s}}{q^k-1}\, \sum_{u \in W_{\omg} \setminus \{0\}} ||\omega_{u}||\eeq
For any $u \in W_{\omg}$, the form $\omg_u \in \wedge^{k-1} (V/\bF u)^*$ is decomposable if and only if $\iota_u \omg$ is decomposable. This easily follows by working with a  basis for $V = W_{\omg} \oplus V_{\omg}$ that extends $u$.
Let $L_{\omg}$ denote the $k+s-1$ dimensional subspace of $\bP(\wedge^{k-1} U_{\omg})$ defined by
\[ L_{\omg} = \bP \{ \iota_u \omg : u  \in W_{\omg}\}\]
Since $\omg$ is indecomposable, it follows by Lemma  \ref{dec_lemma}, that $L_{\omg}$ is not contained in $G_{k-1}(U_{\omg})$. Hence, by Proposition \ref{wt_prop} 
we get
\[|L_{\omg} \cap G_{k-1}(U_{\omg})| \leq 1+q +2 q^2+q^3 + \dots+q^{k+s-2}.\]
Therefore, $|L_{\omg} \setminus G_{k-1}(U_{\omg})| \geq q^{k+s-1} - q^2$. 
On the other hand $|L_{\omg} \setminus G_{k-1}(U_{\omg})| \leq |L_{\omg}|$. Putting these inequalities together we get:
\beq \label{eq:wt2eq2} 
|L_{\omg} \setminus G_{k-1}(U_{\omg})|  = q^{k+s-1} - q^2 + O^+(q^{k+s-2})  = q^{k+s-1} + O(q^{k+s-2}) \eeq
where $O^+(q^m)$ denotes  a positive quantity which is $O(q^m)$. Let 
\[ \delta' := \delta(k-1,n-1) = (k-1)(n-k).\] We note that $q^{\delta'} = q^{\delta}/q^{n-k}$. The weight $||\omg_u||$ equals $q^{\delta'}$  if $\iota_u \omg$ is decomposable, and if $\iota_u \omg$ is indecomposable, then \[ ||\omega_u|| = q^{\delta'}+q^{\delta'-2} +O(q^{\delta'-3}), \] by the inductive hypothesis. Using this in \eqref{eq:wt2eq1}, we get:
\begin{eqnarray*}
\frac{q^s (q^k-1)}{q-1}\, ||\omg|| &=&   (q^{\delta-2} + O(q^{\delta-3}))|L_{\omg} \setminus G_{k-1}(U_{\omg})|  + \; q^{\delta}\, |\bP(W_{\omg})| \\
& =& (q^{\delta-2} + O(q^{\delta-3})) (q^{k+s-1} \!+\! O(q^{k+s-2}))  + \frac{q^{\delta}(q^{k+s}-1)}{q-1}
\end{eqnarray*}

Simplifying this, we get:
\begin{eqnarray} \label{eq:wt2eq3}
\nonumber ||\omg||&=& q^{\delta} + q^{\delta-s}\, \frac{q^s-1}{q^k-1} + (q^{\delta-2} + O(q^{\delta-3})) \,  \frac{q^{k} +  O(q^{k-1})}{q^k-1}\nonumber \\
&=& q^{\delta} + O(q^{\delta-k})+ q^{\delta-2} + O(q^{\delta-3})  \nonumber \\
&=& q^{\delta} +q^{\delta-2} + O(q^{\delta-3}) 
\end{eqnarray}
\eep
\begin{theorem} \label{thm_codim_r}
Let $L$ be a codimension $r$ subspace of $\bP(\wkv)$. If Ann$(L) \subset G_k(V^*)$ then \[||L|| =  q^{\delta} +q^{\delta-1} + \dots + q^{\delta-r+1}.\]
If Ann$(L) \not\subset G_k(V^*)$ and  $r \geq 3$ then:
\[ 
||L|| = q^{\delta} +q^{\delta-1} +2 q^{\delta-2} +O(q^{\delta-3}) \]
If Ann$(L) \not\subset G_k(V^*)$ and  $r= 2$ then:
\[ 
||L|| = q^{\delta} +q^{\delta-1} + q^{\delta-2} + O(q^{\delta-3}) \]
\end{theorem}
\bep When Ann$(L) \subset G_k(V^*)$, the assertion is  a restatement of part 2) of Theorem \ref{thmNogin}. We now assume Ann$(L) \not\subset G_k(V^*)$. 
 Using the formula \eqref{eq:normLH} and Theorem \ref{wt2_thm}, we get:
\begin{eqnarray} \label{eq:normLqrm1}
q^{r-1} \, || L || &=& |\text{Ann}(L) \cap G_k(V^*)| \, q^{\delta}
+ |\text{Ann}(L)\!\setminus\! G_k(V^*)| \, (q^{\delta} +q^{\delta-2} + O(q^{\delta-3})) 
\nonumber \\&=& q^{\delta} \, (1+q+ \dots+q^{r-1})  +  |\text{Ann}(L)\!\setminus\! G_k(V^*)| \,  (q^{\delta-2} + O(q^{\delta-3}))
\nonumber  \end{eqnarray}
Since  Ann$(L) \not\subset G_k(V^*)$ and $r \geq 3$, we use Proposition \ref{wt_prop} to obtain (as in the proof of Theorem \ref{wt2_thm}):
\[  |\text{Ann}(L)\!\setminus\! G_k(V^*)| = q^{r-1} + O(q^{r-2}). \] 
Using this in \eqref{eq:normLqrm1}, we get:
\begin{eqnarray}
|| L || &=& q^{\delta} +q^{\delta-1}+ \dots+q^{\delta-r+1}  +   (q^{\delta-2} + O(q^{\delta-3})) \nonumber\\
&=&  q^{\delta} +q^{\delta-1}+  2 q^{\delta-2} + O(q^{\delta-3})
\nonumber  \end{eqnarray}
In the case when $r=2$,   Ann$(L)$ is a line in $\bP(\wedge^k V^*)$. As mentioned in the proof of Proposition \ref{codim1_prop}, if a line is not contained in $G_k(V^*)$, then it meets
$G_k(V^*)$ in at most two points. Hence $|\text{Ann}(L) \!\setminus\! G_k(V^*)|$ being $q,q-1$ or $q-2$, can be written as  $q +O(1)$. Using this in \eqref{eq:normLqrm1}, we get
\begin{eqnarray*}
|| L || &=&  \frac{1}{q} \left[ q^{\delta}(1 +q)  + (q+O(1)) (q^{\delta-2} + O(q^{\delta-3})) \right] \\
&=& q^{\delta} + q^{\delta-1} +q^{\delta-2} +O(q^{\delta-3}) 
\end{eqnarray*}
\eep

\section{Proof of the main result} \label{proofmain}
We recall from \S  \ref{sec2} the formulas \eqref{eq:E_r} and \eqref{eq:ukncard}
\begin{eqnarray*}
\gamma(k,n) &=&  E_1 - E_2 +E_3- \dots +(-1)^{N-1} E_N \\
E_r &=& \sum_{\{I_1, \dots, I_r\} \in I^r_{k,n}} | C_{I_1} \cup \dots \cup C_{I_r}|
\end{eqnarray*}
From \eqref{eq:CI}, we get  \beq \label{eq:E1} E_1 = N \, q^{\delta} \eeq

If $L$ is the codimension $r$ subspace of $\bP(\wedge^k V)$ defined by
\beq \label{eq:AnnL}  \text{Ann}(L) = \bP \left( \text{Span}(e^{I_1}, \dots, e^{I_r}) \right)\eeq
then  $|C_{I_1} \cup \dots \cup C_{I_r}|=||L||$. 
If $r=2$, the line Ann$(L)$ of $\bP(\wedge^k V^*)$ joining  $e^I$ and $e^J$  is contained in $G_k(V^*)$ if and only if, the $k$-dimensional subspaces of $V^*$ represented by $e^I$ and $e^J$  have a $k-1$ dimensional intersection. In other words there is a  multi-index $K \in I_{k-1,n}$ with $I$ and $J$ of the form $I = K \cup \{i\}$ and $J = K \cup \{j\}$.
Clearly there are  
\[ \textstyle \binom{n}{k-1} \, \binom{n-k+1}{2} = N \delta /2 \]
such elements of $I^2_{k,n}$.
Therefore, by Theorem \ref{thm_codim_r} we get:
\begin{eqnarray} \label{eq:E2}
E_2 &=&  \textstyle \frac{N \delta}{2} \, (q^{\delta} +q^{\delta-1}) + \textstyle \left(\binom{N}{2} - N \delta/2  \right) \, \left(q^{\delta} + q^{\delta-1} +q^{\delta-2} +O(q^{\delta-3}) \right) \nonumber \\
&=& \textstyle \binom{N}{2} \,  (q^{\delta} +q^{\delta-1})  + \frac{N^2 - N - N \delta}{2} \, q^{\delta-2} + O(q^{\delta-3}) \nonumber
  \end{eqnarray}

Now, let $r \geq 3$. The subspace Ann$(L)$ of \eqref{eq:AnnL} is contained in $G_k(V^*)$ if and only if it is contained in a maximal linear subspace $\pi_{\alpha}$ or  $\pi^{\gamma}$ for some $\alpha \in G_{k-1}(V^*)$ or some $\gamma \in G_{k+1}(V^*)$. 
Moreover, these two cases are disjoint, because Ann$(L)$ is $r-1$ dimensional and the intersection of a $\pi_{\alpha}$ and a $\pi^{\gamma}$ is at most one dimensional (part 3) of Fact \ref{fact0}).
Suppose Ann$(L) \subset \pi_{\alpha}$. By part 4) of Fact \ref{fact0}, it follows that 
$\alpha$ is the intersection of the points of $G_k(V^*)$ represented by $e^{I_{\mu}}$ and $e^{I_{\nu}}$ for any pair $1 \leq \mu < \nu \leq r$. In other words, there is a $J \in I_{k-1,n}$  common to all the multi-indices $I_1, \dots, I_r$. Clearly there are 
\beq  \label{eq:c1} 
c_1(r):= \begin{cases} \textstyle \binom{n}{k-1} \, \binom{n-k+1}{r}  = \frac{k N}{n-k+1} \, \binom{n-k+1}{r} &\!\!\text{if  } r \leq n-k+1\\
0 &\!\!\text{if  } r > n-k+1 \end{cases}
\eeq
such elements of $I^r_{k,n}$.

Suppose Ann$(L) \subset \pi^{\gamma}$. By part 4) of Fact \ref{fact0}, it follows that 
 for any pair $1 \leq \mu < \nu \leq r$, $\gamma$ is the vector space sum of the points of $G_k(V^*)$ represented by $e^{I_{\mu}}$ and $e^{I_{\nu}}$. In other words, there is a $\tilde J \in I_{k+1,n}$ containing all the multi-indices $I_1, \dots, I_r$. Clearly there are
 \beq \label{eq:c2} 
c_2(r):= \begin{cases}  \textstyle  \binom{n}{k+1} \, \binom{k+1}{r}  = \frac{N (n-k)}{k+1} \,  \binom{k+1}{r} &\!\!\text{if  } r \leq k+1 \\
0 &\!\!\text{if  } r > k+1 \end{cases}
 \eeq
 such elements of $I^r_{k,n}$. (A derivation of $c_1(r)$ and $c_2(r)$ can also be found in \cite[Corollary 4.4]{GL1}.) Therefore, by Theorem \ref{thm_codim_r} we get: 
\begin{multline*} \sum_{r=3}^N (-1)^{r-1} E_r =    \sum_{r=3}^{N}  \textstyle (-1)^{r-1}\left[(c_1(r)\!+\!c_2(r)) (q^{\delta} \!+\!q^{\delta-1}\!+\!\dots\!+\! q^{\delta-r+1}) \right.\\
+\textstyle \left. (\binom{N}{r} \!-\! c_1(r) \!-\! c_2(r)) ( q^{\delta} +q^{\delta-1} \!+\!2 q^{\delta-2}\!+\!O(q^{\delta-3})) \right]
\end{multline*}
which simplifies to 
\begin{multline*}
\sum_{r=3}^N (-1)^{r-1} E_r =   O(q^{\delta-3}) + (q^{\delta} +q^{\delta-1}) \, \sum_{r=3}^N (-1)^{r-1} \textstyle \binom{N}{r}\\ 
\quad +  q^{\delta-2}  \left[ \tfrac{Nk}{(n-k+1)}  \sum_{r=3}^{n-k+1} \textstyle (-1)^{r}  \binom{n-k+1}{r}  + \tfrac{N (n-k)}{(k+1)} \, \displaystyle \sum_{r=3}^{k+1}  \textstyle (-1)^{r}  \binom{k+1}{r} + \displaystyle 2 \, \sum_{r=3}^N (-1)^{r-1} \textstyle \binom{N}{r}
\right]
 \end{multline*}
Adding $E_1 - E_2$ from formulas \eqref{eq:E1} and \eqref{eq:E2} to the above expression and simplifying  we get:  
\begin{eqnarray*}
\gamma(k,n) &=& O(q^{\delta-3}) +  q^{\delta} \left( \sum_{r=1}^N (-1)^{r-1} \textstyle \binom{N}{r}\right) +  q^{\delta-1} \,  
\left(  \sum_{r=2}^N (-1)^{r-1} \textstyle \binom{N}{r}\right)  \\ 
&&    + q^{\delta-2}  \left[ \tfrac{N \delta}{2(n-k+1)(k+1)} \,  \left( k^2 -n k + n  + 3 \right)  +   2  - \tfrac{5N}{2} +\tfrac{N^2}{2} \right]  \\
&=& q^{\delta}+  (1 - N) \, q^{\delta-1} + a_2(k,n) \, q^{\delta-2} +O(q^{\delta-3})
\end{eqnarray*}
\qed

\section{Conclusion} \label{conc}
Since the problem of determining the number $\gamma(k,n;q)$ of $[n,k]$ $q$-ary MDS codes is very difficult and complicated, we have studied the problem of determining asymptotic formulae in  $q$ for $\gamma(k,n;q)$. We have improved the known formula \eqref{eq:GL1} $\gamma(k,n) = q^{\delta} - (N-1) q^{\delta-1} + O(q^{\delta-2})$ by
by the formula \eqref{eq:asymp}  $\gamma(k,n) = q^{\delta} - (N-1) q^{\delta-1}  + a_2(k,n) q^{\delta-2} + O(q^{\delta -3})$ where $a_2(k,n)$ is given by 
\eqref{eq:unknown}. The main tool is a closer study of cardinalities of linear sections of the Grassmannian (Theorems \ref{wt2_thm} and \ref{thm_codim_r}).
 The problem of improving this to a formula  $\gamma(k,n) = q^{\delta} - (N-1) q^{\delta-1}  + a_2(k,n) q^{\delta-2} + a_3(k,n) q^{\delta -3} + O(q^{\delta -4})$
is more challenging, and will be studied in future work. On a different note, Theorem
\ref{wt2_thm} can be significantly strengthened: In a future work we will show that
for an indecomposable $k$-form $\omg$, we in fact have
\[ ||\omg|| = q^{\delta} +q^{\delta-2} + O^+(q^{\delta-3}) \]
where $O^+(q^{\delta-3})$ is positive quantity which is  $O(q^{\delta-3})$. This enables us to determine some of the as yet unknown higher weights of the Grassmann code $\mC
(k,n)$. 

\bibliographystyle{plain}
\bibliography{refs_mds}
\nocite{}

\end{document}